\documentclass[useAMS,usenatbib]{mn2e}
\usepackage{graphicx}
\usepackage{latexsym}
\usepackage{psfig}
\usepackage{amssymb,amsmath}
\usepackage{subfigure}
\usepackage{caption}

\title{Imprints of feedback in young gasless clusters?}

\author[R. J. Parker \& J. E. Dale]{Richard  J. Parker$^{1}$\thanks{E-mail: rparker@phys.ethz.ch} and James E. Dale$^{2}$  \vspace*{0.1cm}\\
$^{1}$Institute for Astronomy, ETH Z{\"u}rich, Wolfgang-Pauli-Strasse 27, 8093 Z{\"u}rich, Switzerland\\
$^{2}$Excellence Cluster `Universe', Boltzmannstra{\ss}e 2, 85748 Garching, Germany}

\begin{document}

\pagerange{\pageref{firstpage}--\pageref{lastpage}} \pubyear{2013}

\maketitle

\label{firstpage}

\def\mnras{MNRAS}
\def\apj{ApJ}
\def\aj{AJ}
\def\aap{A\&A}
\def\apjl{ApJL}
\def\apjs{ApJS}
\def\araa{ARA\&A}
\def\pasj{PASJ}
 
\begin{abstract}
We present the results of $N$-body simulations in which we take the masses, positions and velocities of sink particles from five pairs of hydrodynamical simulations 
of star formation by \citet{Dale12a,Dale13} and evolve them for a further 10\,Myr. We compare the dynamical evolution of star clusters that formed under the influence of  mass--loss driven by photoionization feedback, to the evolution of clusters that formed without feedback. We 
remove any remaining gas and follow the evolution of structure in the clusters (measured by the $\mathcal{Q}$--parameter), half-mass radius, central density, surface density and the fraction of bound stars.  There is little discernible difference in the evolution of clusters that formed with feedback 
compared to those that formed without. The only clear trend is that all clusters which form without feedback in the hydrodynamical simulations lose any initial structure over 10\,Myr, whereas some of the clusters 
which form with feedback retain structure for the duration of the subsequent $N$-body simulation. This is due to lower initial densities (and hence longer relaxation times) in the clusters from \citet{Dale12a,Dale13} which formed with feedback, which prevents dynamical mixing from erasing substructure. However, several other conditions (such as supervirial initial velocities) 
also preserve substructure, so at a given epoch one would require knowledge of the initial density and virial state of the cluster in order to determine whether star formation in a cluster 
has been strongly influenced by feedback.  
\end{abstract}

 
\section{Introduction}

The conversion of molecular gas to stars results, at least initially, in stars geometrically grouped into clusters and associations \citep[e.g.][]{Lada03,Lada10,Gieles11}. The transition from embedded clusters whose
dynamical behaviour is dominated by gas, to systems devoid of gas, is still not well understood. Stellar feedback, particularly in the form of photoionizing radiation, winds and 
supernovae from OB-type stars, but also photodissociating radiation and jets from low- and intermediate-mass stars, are strong candidates for the agency of gas expulsion and much 
theoretical work has been devoted to their study \citep*[e.g.][]{Whitworth79,Matzner02,Li06,Krumholz12}.

Although the dynamics of gas-free clusters are, by definition, governed by gravity (including tidal interactions with other bodies) and stellar evolution, the prior hydrodynamic phase and the transition 
from that phase to a purely gravitational system are of crucial importance, since together they determine the initial stellar masses, positions and velocities for the subsequent dynamics-dominated evolution 
of the cluster.

In a recent series of papers, \citet*{Dale12a} and \citet*{Dale13} modelled the formation of massive clusters with and without the influence of feedback. 
These authors studied the influence of mass--loss driven by photoionizing radiation from massive stars ($m >$ 20\,M$_\odot$) on the dynamical evolution of their model clouds by comparing each simulation with a corresponding control run in which feedback from the massive stars was switched off.  The 
simulations were evolved for $\sim$3\,Myr, during which time the clusters generally remained substructured, although differences between the runs with and without feedback were generally limited to 
the respective initial mass functions, which were somewhat more top-heavy for the simulations where star formation was not regulated by feedback.
 
Recently, $N$-body simulations of cluster evolution have been used to demonstrate that the initial conditions of star formation can in some cases be discerned from a process of `reverse engineering' -- comparing snapshots from the simulations to observed clusters and re-calibrating the initial conditions of the simulations. 
By assuming realistic initial morphologies for clusters \citep[i.e.\,\,clumpy, or substructured:][]{Cartwright04,Sanchez09} and subvirial velocities \citep{Peretto06,Furesz08}, it is possible to dynamically 
mass-segregate clusters on very short timescales ($< 1$\,Myr) -- the more clumpy the initial conditions, the more likely mass segregation is to occur \citep{Allison10}. 

Additionally, \citet{Allison12} showed that the more subvirial and substructured a cluster is, the more stars are ejected from the cluster, and the larger their velocities are. The initial virial state of a cluster can also be inferred from the amount of substructure 
present. \citet{Parker12d} demonstrate that a subvirial, substructured cluster loses its substructure on very short timescales, whereas a supervirial (expanding) cluster retains substructure in most cases. 

In principle, it may be possible to find differences between clusters that form with feedback and those that form without by examining their long-term dynamical evolution. In this paper, we perform $N$-body simulations in which we take as 
initial conditions the end results of SPH hydrodynamical simulations of embedded clusters from \citet{Dale12a} and \citet{Dale13}. Here, we have chosen the five pairs of feedback and control simulations from these works in which individual 
stars could be resolved. We assert that the remaining gas is expelled instantaneously by supernovae, and treat the resulting clusters as $N$-body problems. 

Earlier work on the impact of instantaneous gas expulsion on clusters suggested that on the removal of background gas potential, clusters quickly become unbound and disperse \citep*[e.g.][]{Tutukov78,Lada84,Goodwin97,Bastian06}. However, 
if a cluster is initially substructured, then the subclusters (which contain the majority of stars) can be gas-poor and the removal of the potential has only a minimal effect on the global dynamics of the cluster \citep{Kruijssen12a}. This is due to 
the stars and gas being decoupled \citep*{Offner09}, and \citet{Smith11} find that the distribution of initial stellar positions and velocities determines whether the star cluster survives, rather than the star formation efficiency\footnote{Another mechanism to limit the effects of gas expulsion is a density--driven star formation efficiency 
gradient in the cluster \citep{Adams00,Parmentier13}, although this mechanism assumes spherical symmetry, which would require clusters to have lost any initial substructure.}.

Here, we 
remove the gas before the $N$-body evolution and rather than examine the effect of removing the gas potential, we determine what influence (if any) the previous gradual gas expulsion by photoionization has on the subsequent (gas free) behaviour of the systems. 
We briefly describe our initial conditions and 
numerical methods in Section 2. The results of the $N$-body calculations follow in Section 3 and our discussion and conclusions follow in Sections 4 and 5.

\section{Initial conditions}

\subsection{Embedded cluster formation}
\label{embed}

In a recent series of papers, \citet{Dale12a} and \citet{Dale13} presented a series of Smoothed Particle Hydrodynamics (SPH) simulations of the evolution of embedded clusters in GMCs with two initial cloud virial ratios of 0.7 and 2.3 (i.e.\,\,bound or unbound), and a range of masses ($10^4$\,M$_\odot$ -- $10^6$\,M$_\odot$) and radii 
between 2.5 and 100\,pc. The initial density profiles of the gas clouds are all close to $\rho \propto r^{-2}$, since the equation of state used is close to being isothermal. We concentrate here on the lower--mass (1--3$\times10^4$\,M$_\odot$) clouds from \citet{Dale12a} and \citet{Dale13}, since these are the simulations in which individual stars could be resolved.

\citet{Dale12a,Dale13} allowed their model clouds to form a few ionizing sources, then split each simulation in two. In one version, feedback from these massive stars is switched on, while in the second (control) run, feedback is left switched off. The formation and evolution of the clusters is then followed for $\sim$3\,Myr, in an attempt to understand the isolated effect of this form of feedback before the detonation of any supernovae. 

Before the onset of feedback, \citet{Dale12a,Dale13} examined the clusters' structure using the $\mathcal{Q}$-parameter \citep{Cartwright04,Cartwright09} and found that all the clusters have some initial substructure, which is retained at the end of the SPH simulations (some clusters become more substructured, others less substructured). This is consistent with the simulations of star formation by \citet{Girichidis11}, in which the majority of the clusters formed in their simulations (from a range of initial cloud density profiles) were substructured \citep[][see their table 3]{Girichidis12}. 

In the simulations by \citet{Dale12a,Dale13}, the differences between the runs with and without feedback were generally limited to the respective initial mass functions, which were somewhat more top-heavy for the feedback-free simulations. The clouds lost a wide range of fractions of their mass during the formation of stars -- from negligible amounts to more than half -- and mass-loss had a largely commensurate influence on the structure and dynamics of the embedded clusters that formed.

In the work we present here, we only use the simulations from \citet{Dale12a,Dale13} in which individual stars could be resolved (down to the lowest possible stellar mass allowed by the resolution limit of the simulation -- 0.5\,M$_\odot$). This gives us a total of five sets of simulations which we will use as the initial conditions of subsequent $N$-body simulations, to compare the effects of dynamical evolution between clusters that formed under the influence of feedback, and those that formed without. In Table~\ref{cluster_props} we summarise these five pairs of SPH simulations, and present the Run ID from \citet{Dale12a,Dale13} -- the prefix `U' indicates an unbound cloud -- as well as the initial physical parameters of the cloud, the number and total mass of stars formed, and the initial and final $\mathcal{Q}$--parameters in the SPH simulations. 

\begin{table*}
\caption[bf]{A summary of the five different pairs of smoothed particle hydrodynamics (SPH) simulations used as the input initial conditions of our $N$-body integrations. The values in the columns are: the simulation number, corresponding Run ID from \citet[][D12]{Dale12a} or \citet[][D13]{Dale13}, 
status of feedback in the SPH simulation (`on' or `off'), the paper reference,  the initial virial ratio of the original clouds $\alpha_{\rm init}^{\rm SPH}$ (to distinguish bound from unbound clouds), the initial radius of the cloud in the SPH simulation ($R_{\rm cloud}$), the initial mass of the cloud ($M_{\rm cloud}$),   the number of stars that have formed at the end of the SPH simulation ($N_{\rm stars}$), the mass of this cluster ($M_{\rm cluster}$), the $\mathcal{Q}$-parameter in the SPH simulation at the time feedback is initiated in the feedback runs ($\mathcal{Q}_{\rm init}^{\rm SPH}$), and the final $\mathcal{Q}$-parameter in the SPH simulations ($\mathcal{Q}_{\rm fin}^{\rm SPH}$).}
\begin{center}
\begin{tabular}{|c|c|c|c|c|c|c|c|c|c|c|}
\hline 
Sim. No. & Run ID & Feedback & Ref.  & $\alpha_{\rm init}^{\rm SPH}$ &$R_{\rm cloud}$ & $M_{\rm cloud}$ & $N_{\rm stars}$ & $M_{\rm cluster}$ & $\mathcal{Q}_{\rm init}^{\rm SPH}$ & $\mathcal{Q}_{\rm fin}^{\rm SPH}$\\
\hline
1(a) & J & On & D12  & 0.7 & 5\,pc & 10 000\,M$_\odot$ & 685 & 2205\,M$_\odot$ &0.53&0.60\\
1(b) & J & Off & D12 & 0.7 & 5\,pc & 10 000\,M$_\odot$ & 578 & 3207\,M$_\odot$ &0.53&0.49\\
\hline
2(a) & I & On & D12 & 0.7 & 10\,pc & 10 000\,M$_\odot$ & 168 & 805\,M$_\odot$ &0.42&0.38\\
2(b) & I & Off & D12 & 0.7 & 10\,pc & 10 000\,M$_\odot$ & 186 & 1270\,M$_\odot$ &0.42&0.72\\
\hline 
3(a) & UF & On & D13 & 2.3 & 10\,pc & 30 000\,M$_\odot$ & 76 & 836\,M$_\odot$ &0.59&0.55\\
3(b) & UF & Off & D13 & 2.3 & 10\,pc & 30 000\,M$_\odot$ & 66 & 1392\,M$_\odot$ &0.59&0.77\\
\hline
4(a) & UP & On & D13 & 2.3 & 2.5\,pc & 10 000\,M$_\odot$ & 346 & 1957\,M$_\odot$ &0.47&0.57\\
4(b) & UP & Off & D13 & 2.3 & 2.5\,pc & 10 000\,M$_\odot$ & 340 & 2718\,M$_\odot$ &0.47&0.49\\
\hline
5(a) & UQ & On & D13 & 2.3 & 5\,pc & 10 000\,M$_\odot$ & 80 & 648\,M$_\odot$ &0.42&0.46\\
5(b) & UQ & Off & D13 & 2.3 & 5\,pc & 10 000\,M$_\odot$ & 48 & 723\,M$_\odot$ &0.42&0.70\\
\hline
\end{tabular}
\end{center}
\label{cluster_props}
\end{table*}

\subsection{$N$-body evolution}

We now take the final states of five pairs of simulations from  \citet{Dale12a} and \citet{Dale13} and assume that the detonation of the first supernova instantaneously removes any remaining gas from both the feedback and non-feedback simulations, and evolve the resulting gas-free systems with an $N$-body code. 
This allows us to gauge the influence of gradual and prolonged ionization-driven mass-loss on the subsequent long-term evolution of the clusters.

As discussed in \citet{Dale12a} and \citet{Dale13}, and Section~\ref{embed}, the lowest mass star which forms in the SPH simulations has a mass of $\sim$0.5\,M$_\odot$. This resolution-limited IMF should not affect the results presented here, as we are looking for differences between the evolution of simulations with the 
same mass resolution.

We evolve the gas-less clusters using the  $4^{\rm th}$ order Hermite-scheme integrator \texttt{kira}  within the Starlab environment \citep[e.g.][]{Zwart99,Zwart01}. We take the masses, positions and velocities of the sink-particles from the SPH simulations and place these directly into the $N$-body integrator. 
In the majority of the SPH runs the stars are in virial equilibrium, or slightly sub-virial, at the end of the simulation (i.e. the initial conditions for the $N$-body integration). However, the initial conditions for simulations UF, UP and UQ were globally unbound, so that one might expect the stars and clusters formed to be in an unbound configuration. 
However, some parts of the globally unbound clouds become bound due to high velocity gas flows colliding and radiating away kinetic energy, which tends to occur in the dense areas of the clouds where most of the stars form. Therefore, in practice, these unbound clouds form stars that are roughly in virial equilibrium (with virial ratios ranging from 0.4 -- 0.7), apart from Run UF (with feedback), which is highly supervirial, with a virial ratio of 1.9. 

In their recent paper, \citet{Moeckel12} used the output of the SPH simulation of star formation by \citet{Bonnell03,Bonnell08} as the input initial conditions for $N$-body simulations to study the subsequent gravitational evolution of the sub-clusters which formed in the SPH simulation. They deliberately scaled the input velocities of 
their sink-particles to be in virial equilibrium. In this work, we will use as a default the original (non-scaled) velocities from the SPH simulations, but we also run the simulations with the stellar velocities scaled to virial equilibrium in order to make a comparison.

The simulation pairs are then evolved for 10\,Myr, without a background gas potential. 

The simulations contain several stars with masses $>~20$\,M$_\odot$ which are likely to evolve over the 10\,Myr duration of the $N$-body integration. For this reason, we use the \texttt{SeBa} stellar evolution package in the Starlab environment 
\citep{Zwart96,Zwart12}, which provides look-up tables for the evolution of stars according to the time dependent mass-radius relations in \citet*{Eggleton89} and \citet{Tout96}. Typically, \texttt{SeBa} updates the evolutionary status of stars on shorter timescales than the timestep in the \texttt{kira} integrator, although for extremely close systems 
a lag of up to one timestep can occur.  

Due to the relatively young end-time of the simulations we do not impose an external galactic tidal field on the simulations. 

\section{Results}
\label{results}

\begin{figure}
  \begin{center}
\setlength{\subfigcapskip}{10pt}
\subfigure{{\includegraphics[trim=35mm 15mm 15mm 20mm, clip, scale=0.5]{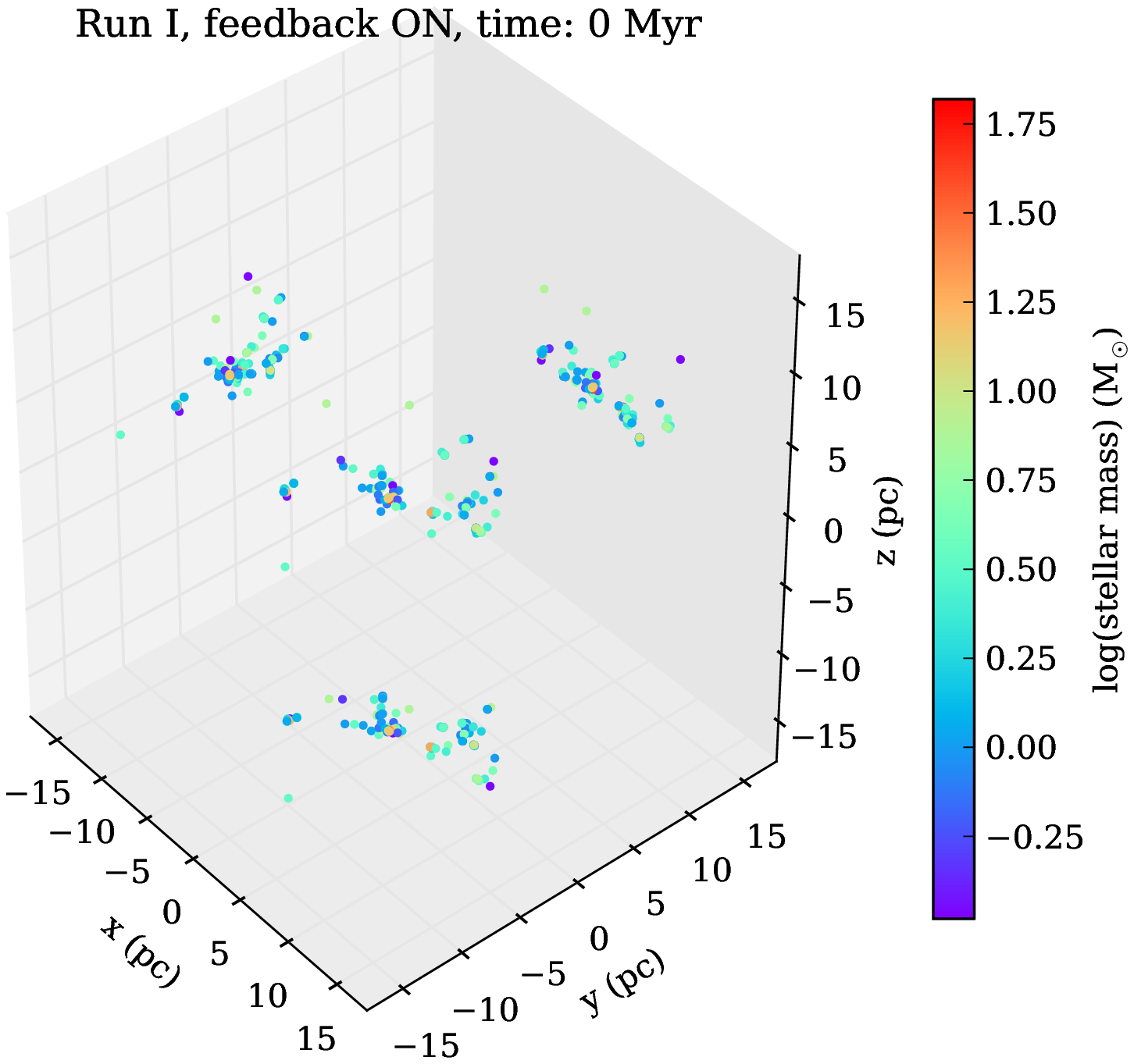}}}
\subfigure{{\includegraphics[trim=35mm 15mm 15mm 20mm, clip, scale=0.5]{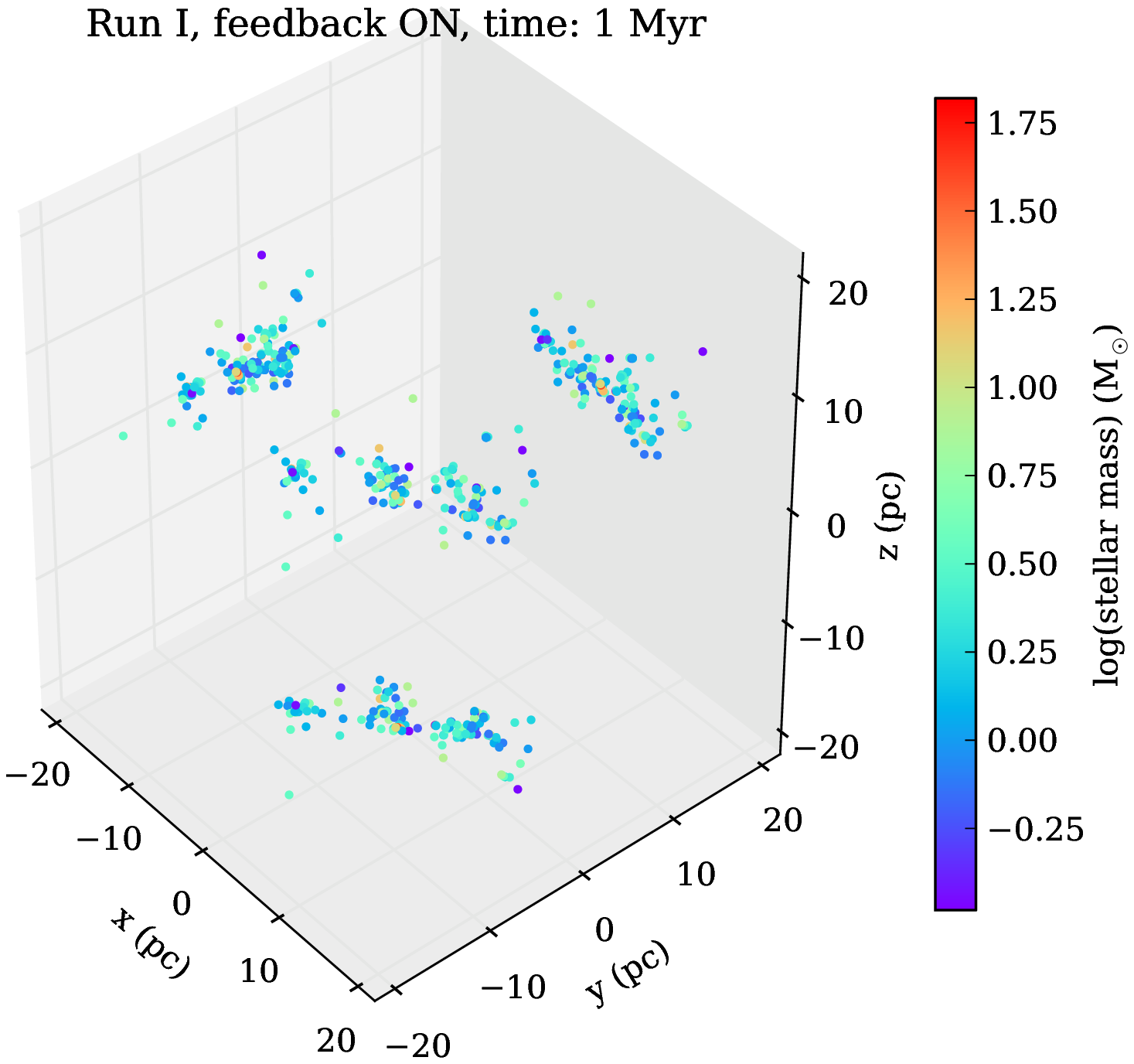}}}
\subfigure{{\includegraphics[trim=35mm 15mm 15mm 20mm, clip, scale=0.5]{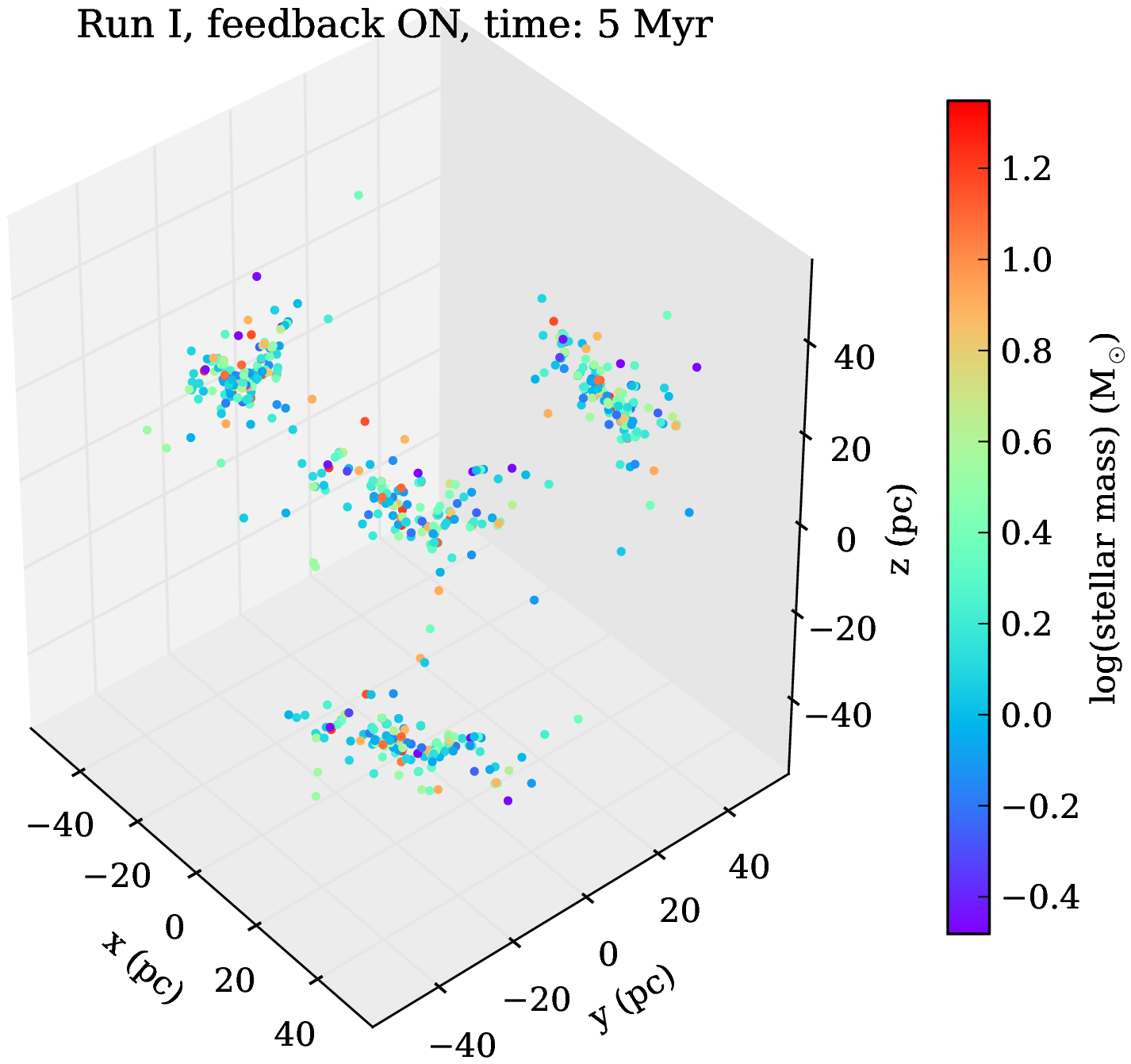}}}
 \end{center}
  \caption[bf]{Evolution of cluster morphology for Run I \emph{with} feedback after 0, 1 and 5\,Myr of $N$-body evolution, with stars colour-coded by mass. Each panel shows a render of the cluster, along with projections of the cluster onto the \emph{x}, \emph{y} and \emph{z} planes. The cluster is still substructured after 5\,Myr.}
  \label{morph_fdbck}
\end{figure}

\begin{figure}
  \begin{center}
\setlength{\subfigcapskip}{10pt}
\subfigure{{\includegraphics[trim=35mm 15mm 15mm 20mm, clip, scale=0.5]{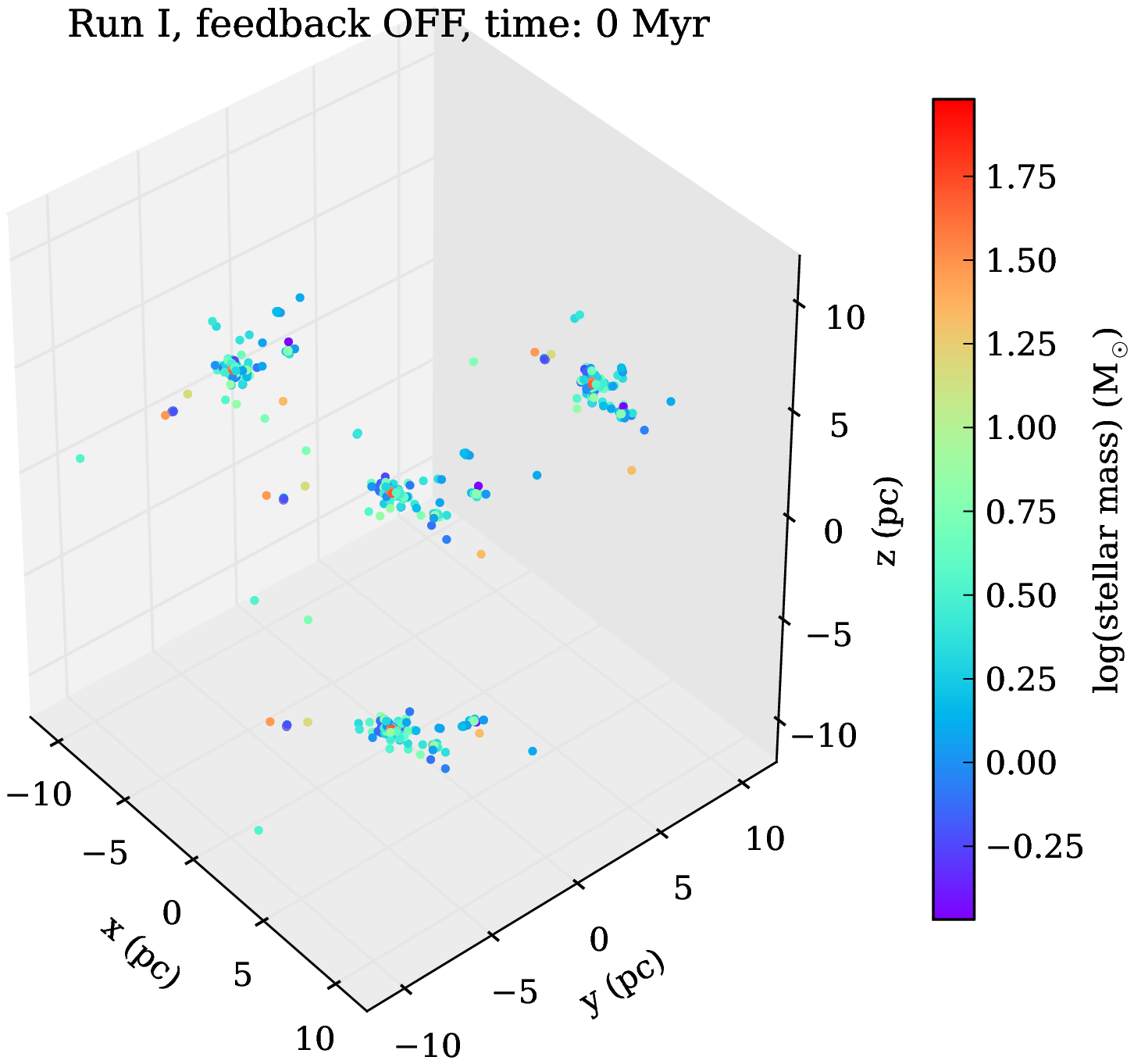}}}
\subfigure{{\includegraphics[trim=35mm 15mm 15mm 20mm, clip, scale=0.5]{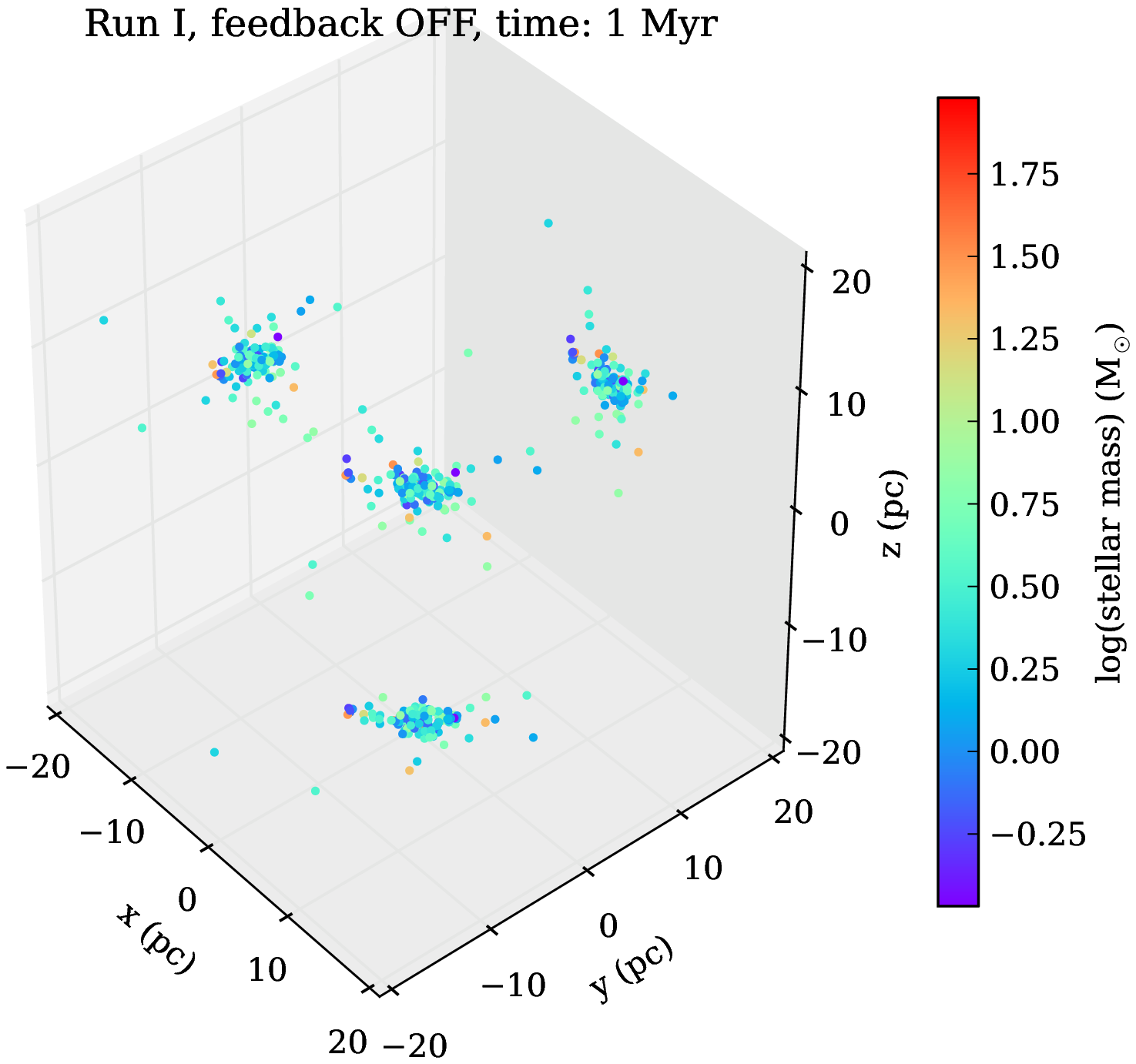}}}
\subfigure{{\includegraphics[trim=35mm 15mm 15mm 20mm, clip, scale=0.5]{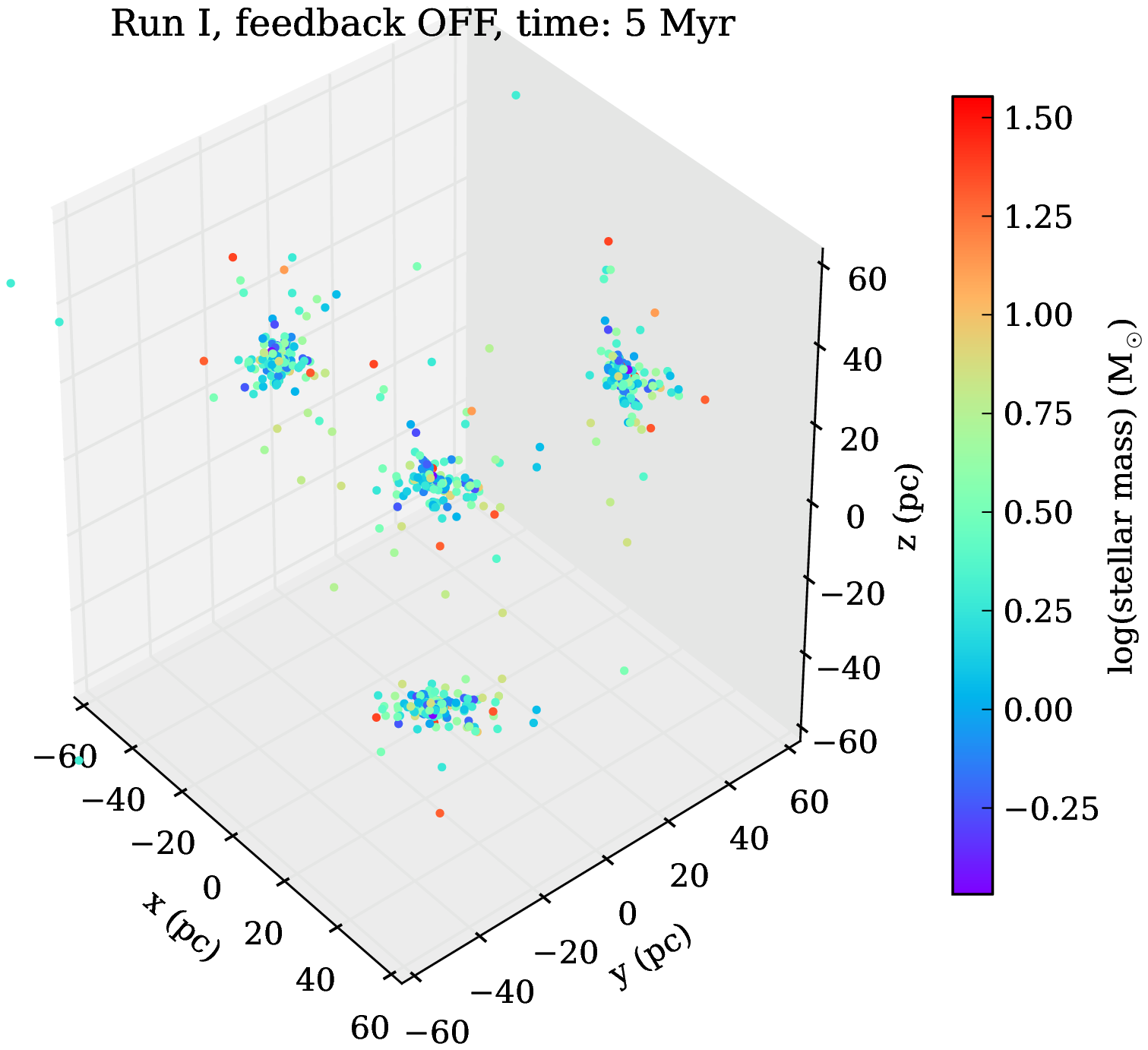}}}
 \end{center}
  \caption[bf]{Evolution of cluster morphology for Run I with \emph{no} feedback after 0, 1 and 5\,Myr of $N$-body evolution, with stars colour-coded by mass. 
 Each panel shows a render of the cluster, along with projections of the cluster onto the \emph{x}, \emph{y} and \emph{z} planes. 
The small amount of initial substructure is erased by dynamics 
and the cluster reaches a smooth, centrally concentrated morphology after 5\,Myr.}
  \label{morph_nofdbck}
\end{figure}

In this Section, we describe the further evolution of the clusters formed in the simulations in \citet{Dale12a} and \citet{Dale13}, using $N$-body simulations. We compare the evolution of cluster structure (as measured by the $\mathcal{Q}$-parameter), mass loss due to stellar evolution, half-mass radii, central volume densities, surface densities, the fraction of bound/unbound stars and search for mass segregation. Our results are necessarily only valid for the initial conditions adopted by \citet{Dale12a,Dale13}. A summary of our results is provided in Table~\ref{cluster_evol}.

\begin{table*}
\caption[bf]{A summary of the results of our $N$-body integrations. The values in the columns are: the simulation number and corresponding Run ID from \citet{Dale12a} or \citet{Dale13} (`a' corresponds to the SPH run with feedback switched on, and  `b' corresponds to the SPH run where feedback was 
switched off), the initial mass of this cluster before $N$-body integration ($M_{\rm cluster, i}$), the final mass after 10\,Myr of $N$-body integration ($M_{\rm cluster, f}$), the initial 
half-mass radius ($r_{1/2,  \rm i}$), the final half-mass radius after 10\,Myr ($r_{1/2,  \rm f}$), initial and final volume densities ($\rho_{\rm i}$ and $\rho_{\rm f}$), initial and final median 
surface densities ($\Sigma_{\rm i}$ and $\Sigma_{\rm f}$) and the initial and final $\mathcal{Q}$-parameters ($\mathcal{Q}_{\rm i}$ and $\mathcal{Q}_{\rm f}$).}
\begin{center}
\hspace*{-0.75cm}\begin{tabular}{|c|c|c|c|c|c|c|c|c|c|c|}
\hline 
Sim. No. & $M_{\rm cluster,  i}$ & $M_{\rm cluster, f}$ & $r_{1/2,  \rm i}$ & $r_{1/2, \rm f}$ & $\rho_{\rm i}$ & $\rho_{\rm f}$ & $\Sigma_{\rm i}$ & $\Sigma_{\rm f}$ & $\mathcal{Q}_{\rm i}$ & $\mathcal{Q}_{\rm f}$ \\
\hline
J, 1(a) & 2205\,M$_\odot$ & 1857\,M$_\odot$ & 1.69\,pc & 5.30\,pc & 54\,M$_\odot$\,pc$^{-3}$  & 1.5\,M$_\odot$\,pc$^{-3}$ & 141\,stars\,pc$^{-2}$ & 2\,stars\,pc$^{-2}$ & 0.60 & 1.89 \\
J, 1(b) & 3207\,M$_\odot$ & 2531\,M$_\odot$ & 1.20\,pc & 15.8\,pc & 220\,M$_\odot$\,pc$^{-3}$  & 0.08\,M$_\odot$\,pc$^{-3}$ & 4518\,stars\,pc$^{-2}$ & 0.4\,stars\,pc$^{-2}$ & 0.49 & 1.91 \\
\hline
I, 2(a) & 805\,M$_\odot$ & 640\,M$_\odot$ & 3.59\,pc & 17.6\,pc & 2.08\,M$_\odot$\,pc$^{-3}$ & 0.01\,M$_\odot$\,pc$^{-3}$  &  83\,stars\,pc$^{-2}$ & 0.1\,stars\,pc$^{-2}$ & 0.38 & 0.79 \\ 
I, 2(b) & 1271\,M$_\odot$ & 751\,M$_\odot$ & 0.70\,pc & 8.72\,pc & 447\,M$_\odot$\,pc$^{-3}$ & 0.14\,M$_\odot$\,pc$^{-3}$ & 102\,stars\,pc$^{-2}$ & 0.3\,stars\,pc$^{-2}$ & 0.72 & 1.39 \\
\hline
UF, 3(a) & 836\,M$_\odot$ & 511\,M$_\odot$ & 6.85\,pc & 32.4\,pc & 0.31\,M$_\odot$\,pc$^{-3}$ & 0.002\,M$_\odot$\,pc$^{-3}$ & 0.6\,stars\,pc$^{-2}$ & 0.01\,stars\,pc$^{-2}$ & 0.55 & 0.74 \\
UF, 3(b) & 1392\,M$_\odot$ & 410\,M$_\odot$ & 1.49\,pc & 20.5\,pc & 50\,M$_\odot$\,pc$^{-3}$  & 0.006\,M$_\odot$\,pc$^{-3}$  & 6\,stars\,pc$^{-2}$ &  0.01\,stars\,pc$^{-2}$ & 0.77 & 1.01 \\
\hline
UP, 4(a) & 1957\,M$_\odot$ & 1587\,M$_\odot$ & 3.85\,pc & 12.4\,pc & 4.09\,M$_\odot$\,pc$^{-3}$  & 0.09\,M$_\odot$\,pc$^{-3}$  & 24\,stars\,pc$^{-2}$ & 0.2\,stars\,pc$^{-2}$ & 0.57 & 1.27 \\
UP, 4(b) & 2718\,M$_\odot$ & 1765\,M$_\odot$ & 3.87\,pc & 19.6\,pc & 5.53\,M$_\odot$\,pc$^{-3}$  & 0.03\,M$_\odot$\,pc$^{-3}$  & 250\,stars\,pc$^{-2}$ & 0.2\,stars\,pc$^{-2}$ & 0.49 & 1.40 \\
\hline
UQ, 5(a) & 648\,M$_\odot$ & 485\,M$_\odot$ & 5.13\,pc & 25.7\,pc & 0.57\,M$_\odot$\,pc$^{-3}$  & 0.003\,M$_\odot$\,pc$^{-3}$  & 2\,stars\,pc$^{-2}$ & 0.04\,stars\,pc$^{-2}$ & 0.46 & 0.72 \\
UQ, 5(b) & 723\,M$_\odot$ & 337\,M$_\odot$ & 1.32\,pc & 13.7\,pc & 37.3\,M$_\odot$\,pc$^{-3}$  & 0.02\,M$_\odot$\,pc$^{-3}$  & 6\,stars\,pc$^{-2}$ & 0.02\,stars\,pc$^{-2}$ & 0.70 & 0.93 \\
\hline
\end{tabular}
\end{center}
\label{cluster_evol}
\end{table*}

\subsection{Cluster morphologies and initial structure}
\label{morph}

In Figs.~\ref{morph_fdbck}~and~\ref{morph_nofdbck} we show the morphologies of Run I with and without feedback, respectively. From top to bottom, we show the initial morphology before $N$-body 
evolution (0\,Myr) and after 1\,Myr and 5\,Myr. Each panel shows a render of the cluster, along with projections of the cluster onto the \emph{x}, \emph{y} and \emph{z} planes.  The simulation with feedback switched on (Fig.~\ref{morph_fdbck}) exhibits a high degree of substructure, which is still evident after 1\,Myr (middle panel) 
and to a lesser degree after 5\,Myr. On the other hand, the simulation without feedback (Fig.~\ref{morph_nofdbck}) appears to be slightly substructured initially, before assuming a smooth, centrally concentrated morphology. 

Whilst Run~I is typical of the differences between all pairs of simulations, it is useful to compare the initial morphologies for the feedback versus no-feedback initial conditions for all simulations. In order to quantify structure, 
we adopt the $\mathcal{Q}$-parameter \citep{Cartwright04,Cartwright09}, which compares the mean length of a minimum spanning tree (MST) joining all the stars in a cluster, $\bar{m}$, to the average separation between stars in the cluster, $\bar{s}$:
\begin{equation}
\mathcal{Q} = \frac{\bar{m}}{\bar{s}}.
\end{equation}
In two dimensions, a cluster is substructured if the $\mathcal{Q}$-parameter is less than 0.8 (lower values correspond to a higher degree of structure), or centrally concentrated if the $\mathcal{Q}$-parameter is greater than 0.8 (higher values correspond to 
a more centrally concentrated morphology). In all cases, we project our clusters along the z--axis, but the results are robust against other projections.

For Run~J, the $\mathcal{Q}$-parameter is 0.60 at the start of the $N$-body integration for the simulation with feedback, and is 0.49 for the simulation with no feedback. On the other hand (and as is visible by comparing the top panels of 
Figs.~\ref{morph_fdbck}~and~\ref{morph_nofdbck}), for Run~I the simulation with feedback is initially more structured, with a $\mathcal{Q}$-parameter of 0.38, as opposed to a $\mathcal{Q}$-parameter of 0.72 for the run with no feedback. 
In the remaining simulations, Run~UF has $\mathcal{Q}$-parameters of 0.55 (with feedback) and 0.77 (no feedback); Run~UP has $\mathcal{Q}$-parameters of 0.57 (with feedback) and 0.49 (no feedback); and Run~UQ has $\mathcal{Q}$-parameters of 
0.46 (with feedback) and 0.70 (no feedback). 

In summary, there is no clear trend for clusters with feedback to be more or less structured than those without at the end of the SPH simulations (and before $N$-body integration).

\subsection{Evolution of cluster structure}

\begin{figure}
\begin{center}
\rotatebox{270}{\includegraphics[scale=0.4]{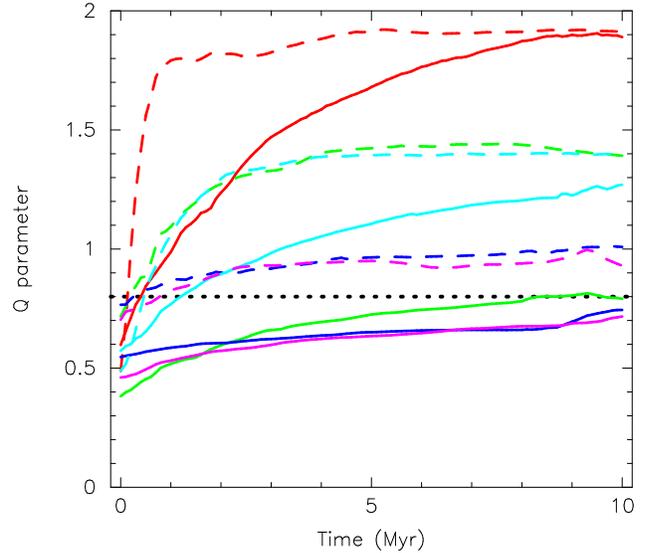}}
\end{center}
\caption[bf]{Evolution of the $\mathcal{Q}$-parameter \citep{Cartwright04} for all simulations. The boundary between centrally concentrated, radially smooth clusters ($\mathcal{Q} > 0.8$) and substructured clusters ($\mathcal{Q} < 0.8$) is shown by the dotted line. 
Simulations \emph{with} feedback are shown by the solid lines; simulations with no feedback are shown by the dashed lines. The red lines are Runs J, green lines are Runs I, dark blue lines are Runs UF, cyan lines are Runs UP and the magenta lines are Runs UQ. Runs with feedback preserve structure for longer as the cluster evolves, and three simulations with feedback (I, UF and UQ) remain substructured beyond 5\,Myr.}
\label{Qpar_all}
\end{figure}

We now examine the evolution of the structure of the model clusters over the duration of the $N$-body integration, using the $\mathcal{Q}$-parameter \citep{Cartwright04}. In Fig.~\ref{Qpar_all} we compare the evolution of the $\mathcal{Q}$-parameter with time in the five pairs of simulations. 
The simulations that formed with feedback are shown by the solid lines, and the simulations that formed without feedback are shown by the dashed lines. The colours correspond to the following simulations; red--Run~J, green--Run~I, dark blue--Run~UF, cyan--Run~UP and magenta--Run~UQ. 
As detailed in Section~\ref{morph}, all clusters which form in the simulations of \citet{Dale12a,Dale13} have substructure initially (i.e.\,\,$\mathcal{Q} < 0.8$), but there is no trend towards a particular regime of $\mathcal{Q}$ as a consequence of feedback being switched on or off.  

In all cases, the clusters lose substructure as they evolve, so that their $\mathcal{Q}$-parameters increase. The clusters resulting from simulations without feedback (the dashed lines in Fig.~\ref{Qpar_all}) lose substructure more quickly, and all of these no--feedback runs attain a centrally 
concentrated morphology with final $\mathcal{Q}$-parameters of 1.91 (Run~J), 1.39 (Run~I), 1.01 (Run~UF), 1.40 (Run~UP) and 0.93 (Run~UQ). 

Conversely, the clusters resulting from simulations \emph{with} feedback lose their structure more slowly, and two simulations (Run~UF and Run~UQ) remain substructured even after 10\,Myr of $N$-body integration (the solid blue and magenta lines in Fig.~\ref{Qpar_all}), with final $\mathcal{Q}$-parameters of 
0.74 and 0.72, respectively. Run~I with feedback (the solid green line) retains substructure for 8.3\,Myr, with a final $\mathcal{Q}$-parameter of 0.8 (i.e.\,\,straddling the boundary between substructured and centrally concentrated). Runs~J and UP with feedback (the solid red and cyan lines) lose their substructure within the first 2\,Myr of $N$-body integration, with final 
$\mathcal{Q}$-parameters of 1.89 and  1.27, respectively.

Not only do the clusters resulting from simulations with feedback lose structure more slowly, but in all pairs of simulations the $\mathcal{Q}$-parameters are systematically lower after $N$-body evolution for the runs \emph{with} feedback compared to those without. After 10\,Myr of $N$-body integration this difference is 
most pronounced for Run~I (the green lines), with a difference 0.59 in $\mathcal{Q}$-parameter, and least pronounced for Run~J (the red lines), with a difference 0.02 in $\mathcal{Q}$-parameter (though this simulation displays a difference of order 0.7 in $\mathcal{Q}$-parameter earlier in the $N$-body integration at $\sim$0.5\,Myr).

\subsection{Stellar evolution}

\begin{figure*}
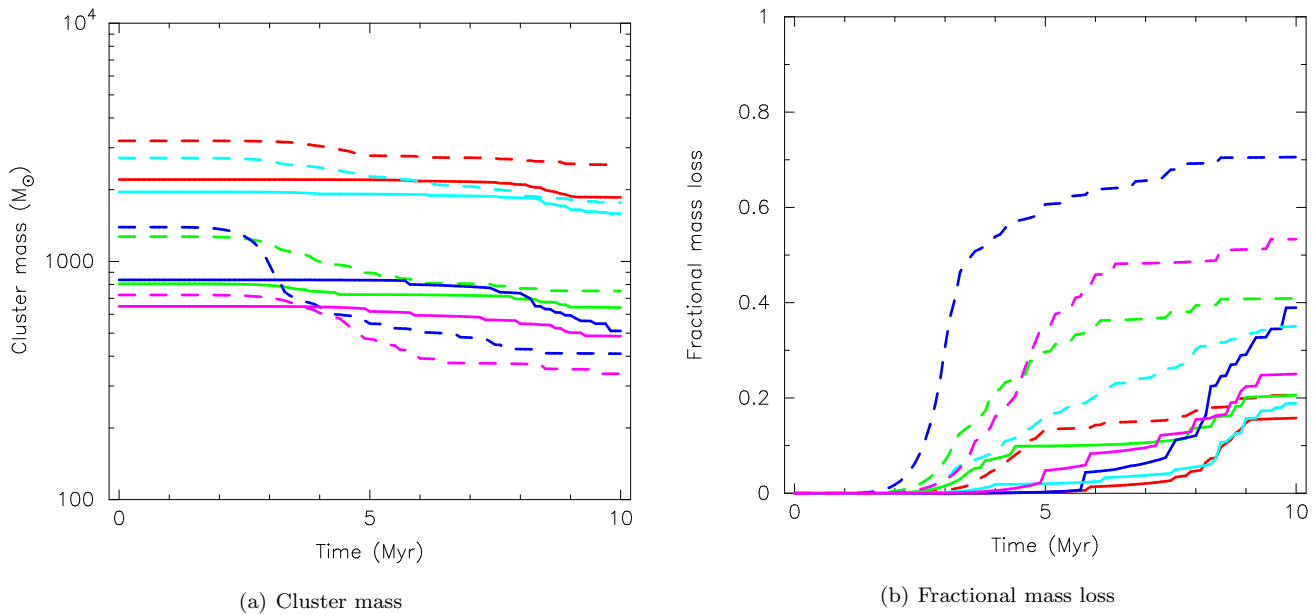

  \begin{center}
\setlength{\subfigcapskip}{10pt}
\hspace*{-0.3cm}
\subfigure[Cluster mass]{\label{stellar_ev-a}\rotatebox{270}{\includegraphics[scale=0.4]{Mass_loss_all_Ovel_stellar_ev.ps}}}  
\hspace*{0.5cm}
\subfigure[Fractional mass loss]{\label{stellar_ev-b}\rotatebox{270}{\includegraphics[scale=0.4]{Mass_frac_all_Ovel_stellar_ev.ps}}}
\end{center}
  \caption[bf]{The effects of stellar evolution on the total mass within each simulation. In panel (a) we show the evolution of the total cluster mass and in panel (b) we show the fraction of the initial mass of the cluster which is lost through 
stellar evolution (we include all stars, irrespective of whether or not they are still bound). Simulations \emph{with} feedback are shown by the solid lines; simulations with no feedback are shown by the dashed lines. The red lines are Runs J, green lines are Runs I, dark blue lines are Runs UF, cyan lines are Runs UP and the magenta lines are Runs UQ. 
The simulations with no feedback have a more top-heavy IMF, and lose more mass through stellar evolution (more than 50\,per cent of their initial mass in simulations UF and UQ).}
  \label{stellar_ev}
\end{figure*}

We show the effects of stellar evolution on our $N$-body simulations\footnote{We also performed the same suite of $N$-body simulations with stellar evolution switched off, and the results are qualitatively similar (such as the evolution of the $\mathcal{Q}$-parameter, evolution of the half-mass radius 
and evolution of surface density).} in Fig.~\ref{stellar_ev}. 
The simulations of clusters with feedback are shown by the solid lines, whereas the clusters without feedback are shown by the dashed lines.  The colours correspond to the following simulations; red--Run~J, green--Run~I, dark blue--Run~UF, cyan--Run~UP and magenta--Run~UQ. 
In Fig.~\ref{stellar_ev-a} we show the evolution of total cluster mass over the duration of the $N$-body integration, and in Fig.~\ref{stellar_ev-b} we show the fraction of the initial stellar mass that remains, over the 10\,Myr of $N$-body integration. 

The simulations of clusters that formed with feedback have lower initial masses than those that formed without. This is because in the simulations without feedback, star formation is not suppressed by feedback mechanisms, and this can lead to the growth of relatively massive stars, which in turn results in a top-heavy IMF in the clusters 
without feedback \citep{Dale12a,Dale13}. However, the formation of the more massive stars in the clusters without feedback means that a higher fraction of stars lose a significant amount of mass (and go supernova) within the duration of the $N$-body integration. This is most apparent in Runs~UF and UQ (the dashed dark blue and magenta lines 
in Fig.~\ref{stellar_ev-b}, respectively), which both lose more than 50\,per cent of their initial mass. In contrast, the clusters that formed \emph{with} feedback (the solid lines in Fig.~\ref{stellar_ev-b}) do not contain as many massive stars, and typically lose only $\sim$\,20\,per cent of their original mass.

\subsection{Half-mass radii}
\label{half-mass}

\begin{figure}
\begin{center}
\rotatebox{270}{\includegraphics[scale=0.4]{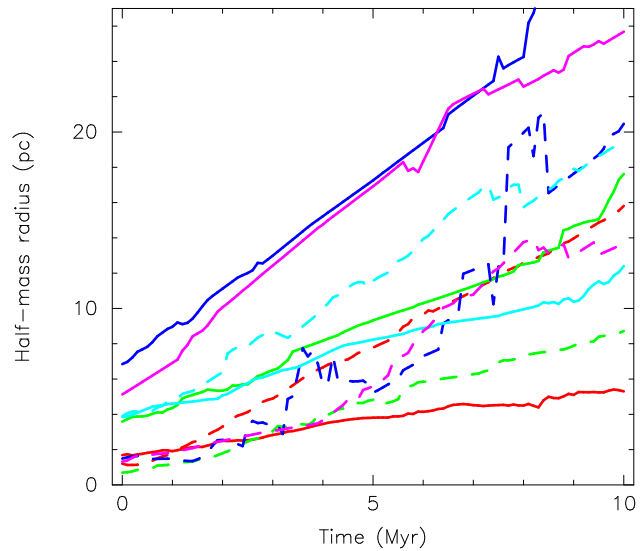}}
\end{center}
\caption[bf]{Evolution of the half-mass radius for all simulations. Simulations \emph{with} feedback are shown by the solid lines; simulations with no feedback are shown by the dashed lines. The red lines are Runs J, green lines are Runs I, dark blue lines are Runs 
UF, cyan lines are Runs UP and the magenta lines are Runs UQ. There is no clear dependence of the evolution of half-mass radius on initial conditions (i.e.\,\,feedback versus no feedback).}
\label{Rhm_all}
\end{figure}

In Fig.~\ref{Rhm_all}, we compare the evolution of the half--mass radii in the simulation pairs. We define the half-mass radius based on the centre of mass for the cluster, which is problematic if the cluster is initially substructured \citep*[e.g.][]{Parker11c}. As most of the clusters eventually 
attain a centrally concentrated morphology we will examine the differences between simulations{\bf,} but with this caveat in mind. (A better measure of how much the cluster expands is perhaps to compare the cumulative distributions of stellar surface densities, as they do not depend on morphology \citep[e.g.][]{Parker12d},  
which we will explore in Section~\ref{surface_dens}.) 

There is no obvious pattern in behaviour of the half-mass radii. In Run~J (the red lines), the feedback-influenced cloud loses a substantial ($\sim$~20\,per cent) fraction of its gas to feedback before the assumed supernova detonation (and the subsequent $N$-body evolution), 
but the resulting cluster nevertheless expands more slowly than does the one from the corresponding run without feedback. The initial half-mass radii for this simulation pair are 1.69\,pc for the cluster that formed with feedback, and 1.20\,pc for the cluster that formed without. However, after 10\,Myr 
the cluster that formed with feedback has a half-mass radius of 5.3\,pc compared to 15.8\,pc for the cluster that formed without.  A similar result occurs for Run~UP (the cyan lines); the initial half-mass radii are 3.85\,pc (for the cluster that formed with feedback) and 3.87\,pc (for the cluster that formed 
without feedback). The cluster with feedback (the solid cyan line) then expands more slowly and has a final half-mass radius of 12.4\,pc compared to 19.6\,pc for the cluster without feedback (the dashed cyan line). 

For Run~I, the cluster that formed with feedback has a much larger half-mass radius (3.59\,pc) than the simulation of the cluster that formed without feedback (0.70\,pc) and both clusters expand at a similar rate (the green lines in Fig.~\ref{Rhm_all}). However,  the high degree of substructure present in 
the simulation with feedback suggests that the determination of half-mass radius (which assumes a symmetric, centrally concentrated morphology) could be misleading.  Similarly, in Runs~UF~and~UQ (the dark blue and magenta lines) the simulations that formed with feedback are highly substructured and the small numbers of stars 
in these simulations results in the evolution of the half-mass radius being rather noisy. This is in part due to the occurrence of supernovae (compare the dips in half-mass radius with the mass-loss in Fig.~\ref{stellar_ev-b}), which redefines the `half-mass' on very short timescales. In contrast to Run~I and Run~UP, the 
half-mass radii for the 3 other simulations is lower for those clusters that formed without feedback, compared to those that formed with feedback.

\subsection{Central densities}
\label{cent_dens}

\begin{figure}
\begin{center}
\rotatebox{270}{\includegraphics[scale=0.4]{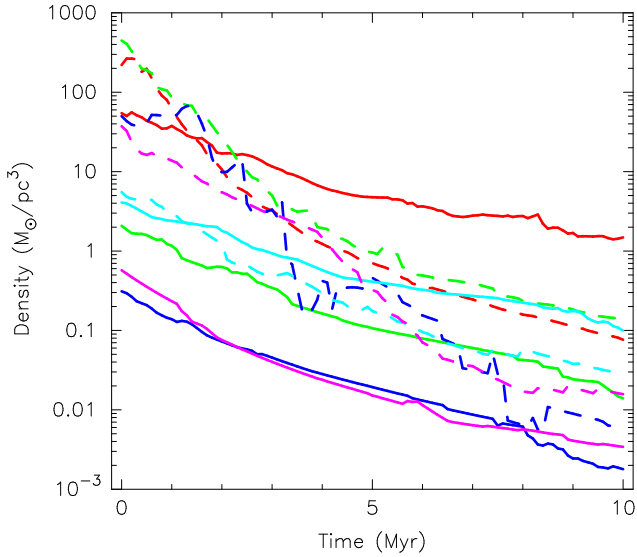}}
\end{center}
\caption[bf]{Evolution of the central stellar density (i.e.\,\,the density within the half-mass radius) for all simulations. Simulations \emph{with} feedback are shown by the solid lines; simulations with no feedback are shown by the dashed lines. The red lines are Runs J, green lines are Runs I, dark blue lines are Runs 
UF, cyan lines are Runs UP and the magenta lines are Runs UQ. Simulations that formed with feedback all have lower initial densities than those that form without feedback, but the subsequent evolution is non-uniform.}
\label{Dens_all}
\end{figure}

Although ill-defined for substructured clusters, we use the half-mass radius to determine a central density, i.e.\,\,the stellar mass density within the half-mass radius. In Fig.~\ref{Dens_all} we plot the evolution of this density for all five pairs of simulations. The evolution of simulations that formed with feedback is shown by the solid lines, 
and the evolution for clusters that formed without feedback is shown by the dashed lines. It is apparent that the initial density is lower for clusters that formed with feedback in the \citet{Dale12a,Dale13} simulations, than for those that formed without. In the most extreme case, Run~I that formed with feedback (the solid green line in Fig.~\ref{Dens_all}) has an initial 
density of 2.1\,M$_\odot$\,pc$^{-3}$, compared to the cluster that formed without feedback (the dashed green line), which has an initial density of 447\,M$_\odot$\,pc$^{-3}$. Run UF (the blue lines) also exhibits a large difference, with an initial density of 0.3\,M$_\odot$\,pc$^{-3}$ (with feedback) compared to 49.9\,M$_\odot$\,pc$^{-3}$ 
(without feedback). 

The subsequent evolution of the densities is non-uniform. The clusters that formed without feedback tend to become less dense on faster timescales than those that formed with feedback, due to rapid mass-loss from stellar winds and supernovae in the $N$-body evolution; this is most apparent for Runs~UF~and~UQ (the dashed blue and 
magenta lines), which also lose the most stellar mass over 10\,Myr (see Fig.~\ref{stellar_ev-b}).

\subsection{Surface densities}
\label{surface_dens}

As discussed in Sections~\ref{half-mass}~and~\ref{cent_dens}, determining the half-mass radius in a substructured cluster is problematic, and a more informative measure of cluster expansion may be the evolution of local stellar surface density. We define surface density, $\Sigma$, based on the definition in \citet{Casertano85}:
\begin{equation}
\Sigma = \frac{N - 1}{\pi D^2_N},
\end{equation}
where $N$ is the $N^{\rm th}$ nearest neighbour and $D_N$ is the projected distance to that nearest neighbour. We adopt $N = 7$, in order to detect localised pockets of high stellar density but not be biased by binary or higher order multiple systems. 

\begin{figure*}
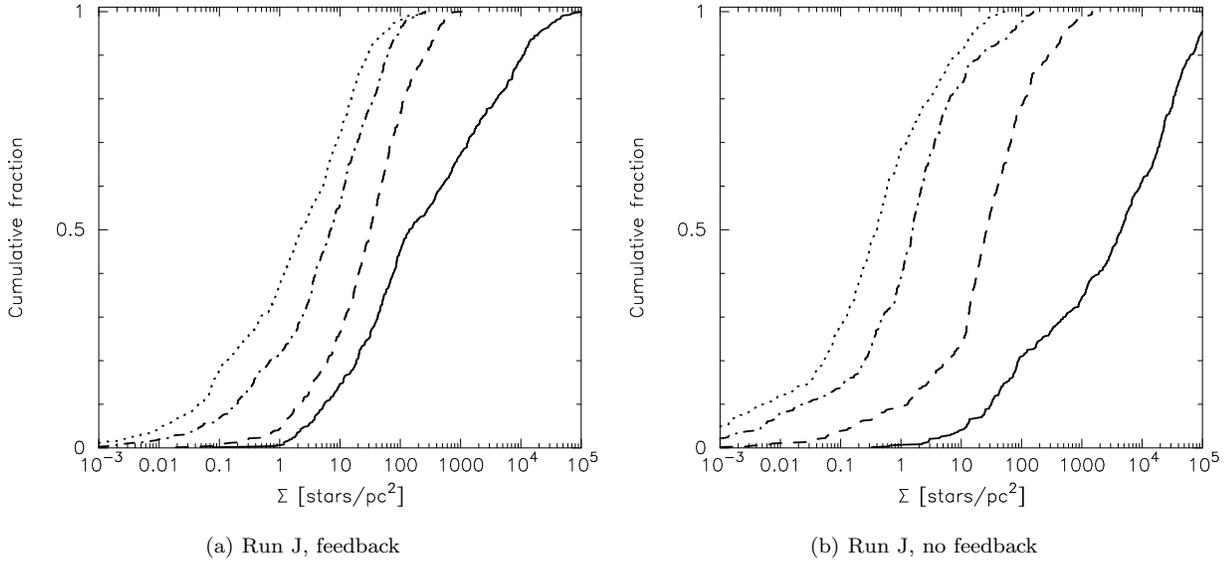

  \begin{center}
\setlength{\subfigcapskip}{10pt}
\hspace*{-0.3cm}
\subfigure[Run J, feedback]{\label{sigma_1_fdbk}\rotatebox{270}{\includegraphics[scale=0.37]{Sigma_Trig_1_Ovel_SEba.ps}}}  
\hspace*{0.2cm}
\subfigure[Run J, no feedback]{\label{sigma_1Nfdbk}\rotatebox{270}{\includegraphics[scale=0.37]{Sigma_TrigN1_Ovel_SEba.ps}}}
\end{center}
  \caption[bf]{Evolution of stellar surface density for Run J (a) with feedback and (b) no feedback. The cumulative distributions of surface density at 0\,Myr (the solid line), 1\,Myr (the dashed line), 5\,Myr (the dot--dashed line) and 10\,Myr (the dotted line) are shown. Before dynamical 
evolution, the simulation with feedback (a) has a median stellar surface density of 141\,stars\,pc$^{-2}$, which decreases to 32\,stars\,pc$^{-2}$ after 1\,Myr, 7\,stars\,pc$^{-2}$ after 5\,Myr and 2\,stars\,pc$^{-2}$ after 10\,Myr. In contrast, the simulation with feedback (b) has an initial median 
stellar surface density of 4518\,stars\,pc$^{-2}$, which decreases to 26\,stars\,pc$^{-2}$ after 1\,Myr, 2\,stars\,pc$^{-2}$ after 5\,Myr and 0.4\,stars\,pc$^{-2}$ after 10\,Myr.}
  \label{sigma_1}
\end{figure*}

In Fig.~\ref{sigma_1} we show plots depicting the cumulative distributions of surface densities for stars at 0, 1, 5 and 10 Myr (the solid, dashed, dot--dashed and dotted lines, respectively) in the $N$-body integration of simulation Run~J. In Fig.~\ref{sigma_1_fdbk} we 
show the evolution of the $\Sigma$~distribution for the cluster that formed with feedback, and in Fig.~\ref{sigma_1Nfdbk} we show the evolution for the cluster that formed without feedback. The cluster that forms with feedback has an initial median surface density 
$\Sigma = 141$\,stars\,pc$^{-2}$, which decreases to 32\,stars\,pc$^{-2}$ after 1\,Myr, 7\,stars\,pc$^{-2}$ after 5\,Myr and 2\,stars\,pc$^{-2}$ after 10\,Myr. In contrast, the cluster that formed with no feedback has a much higher initial surface density, 
4518\,stars\,pc$^{-2}$, which decreases to 26\,stars\,pc$^{-2}$ after 1\,Myr, 2\,stars\,pc$^{-2}$ after 5\,Myr and 0.4\,stars\,pc$^{-2}$ after 10\,Myr. The clusters start with very different surface densities, but converge to similar distributions over the 10\,Myr of 
$N$-body evolution.

\begin{figure*}
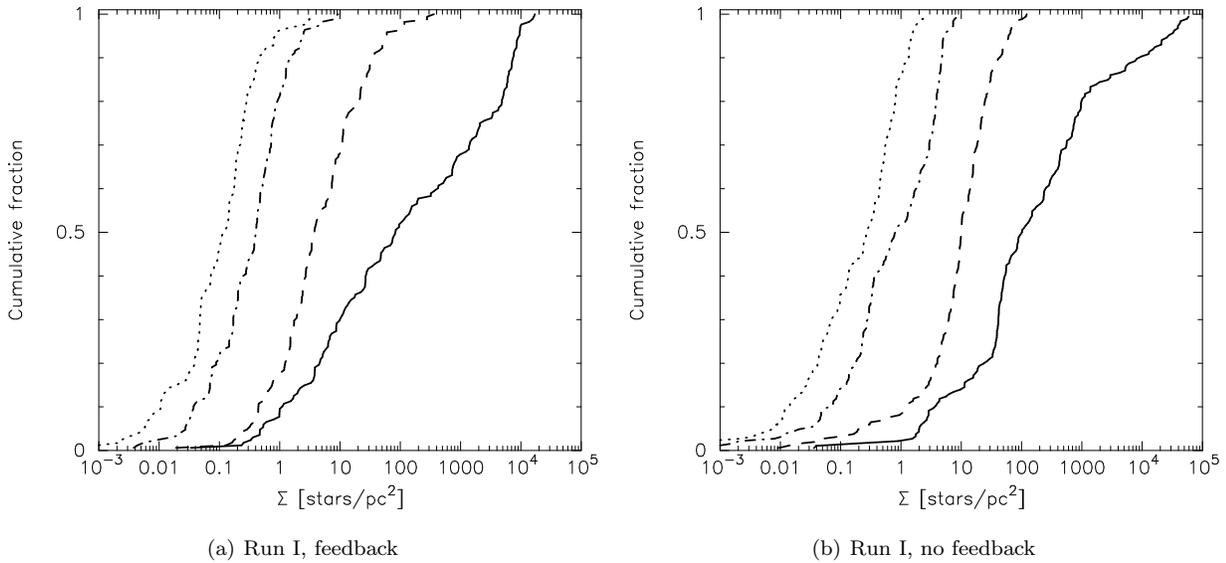

  \begin{center}
\setlength{\subfigcapskip}{10pt}
\hspace*{-0.3cm}
\subfigure[Run I, feedback]{\label{sigma_2_fdbk}\rotatebox{270}{\includegraphics[scale=0.37]{Sigma_Trig_2_Ovel_SEba.ps}}}  
\hspace*{0.2cm}
\subfigure[Run I, no feedback]{\label{sigma_2Nfdbk}\rotatebox{270}{\includegraphics[scale=0.37]{Sigma_TrigN2_Ovel_SEba.ps}}}
\end{center}
  \caption[bf]{Evolution of stellar surface density for Run I (a) with feedback and (b) no feedback. The cumulative distributions of surface density at 0\,Myr (the solid line), 1\,Myr (the dashed line), 5\,Myr (the dot--dashed line) and 10\,Myr (the dotted line) are shown. Before dynamical 
evolution, the simulation with feedback (a) has a median stellar surface density of 83\,stars\,pc$^{-2}$, which decreases to 4\,stars\,pc$^{-2}$ after 1\,Myr, 0.4\,stars\,pc$^{-2}$ after 5\,Myr and 0.1\,stars\,pc$^{-2}$ after 10\,Myr. The simulation with feedback (b) has a similar initial median 
stellar surface density of 102\,stars\,pc$^{-2}$, which decreases to 10\,stars\,pc$^{-2}$ after 1\,Myr, 0.8\,stars\,pc$^{-2}$ after 5\,Myr and 0.3\,stars\,pc$^{-2}$ after 10\,Myr.}
  \label{sigma_2}
\end{figure*}

From inspection of the evolution of the $\Sigma$ distribution for Run~J, it appears that the cluster that formed without feedback has been able to form with a much higher surface density than the simulation with feedback, as suggested by the volume density measurements in Section~\ref{cent_dens}. 
However, in Fig.~\ref{sigma_2} we show the initial surface densities for the simulation Run~I with feedback (panel a) and without feedback (panel b) by the solid lines. Recall that the volume density for this cluster was 2.1\,M$_\odot$\,pc$^{-3}$ for the cluster that formed with feedback, compared to 
447\,M$_\odot$\,pc$^{-3}$  for the simulation that formed without. However, the median surface densities are very similar; the simulation that formed with feedback has $\Sigma = 83$\,stars\,pc$^{-2}$ whereas the simulation that formed without has $\Sigma = 102$\,stars\,pc$^{-2}$. This is due to 
the initially high level of substructure ($\mathcal{Q} = 0.38$) in the simulation, which presents itself as high initial local surface densities (the solid line in Fig.~\ref{sigma_2_fdbk}), but as a low volume density because the volume density is defined from the ill-defined cluster `centre'.

In Fig.~\ref{sigma_2_fdbk} we show the evolution of the $\Sigma$~distribution for the cluster that formed with feedback, and in Fig.~\ref{sigma_2Nfdbk} we show the evolution for the cluster that formed without feedback.  The cluster that forms with feedback has an initial median surface density 
$\Sigma = 83$\,stars\,pc$^{-2}$ (the solid line),  which decreases to 4\,stars\,pc$^{-2}$ after 1\,Myr, 0.4\,stars\,pc$^{-2}$ after 5\,Myr and 0.1\,stars\,pc$^{-2}$ after 10\,Myr (the dashed, dot--dashed and dotted lines, respectively). The cluster that forms without feedback  has a similar initial median 
stellar surface density of 102\,stars\,pc$^{-2}$, which decreases to 10\,stars\,pc$^{-2}$ after 1\,Myr, 0.8\,stars\,pc$^{-2}$ after 5\,Myr and 0.3\,stars\,pc$^{-2}$ after 10\,Myr. 

The remaining  three pairs of simulations also do not show any clear dependence of the evolution of stellar surface density on whether the clusters formed with or without feedback. In all pairs of simulations, the clusters that formed without feedback have higher surface densities (although as we have seen in Fig.~\ref{sigma_2} 
the difference can be marginal), suggesting that the presence of feedback decreases the density of star formation. However, in all simulations, the median surface density decreases significantly (often by 3 -- 4 orders of magnitude) 
over 10\,Myr of $N$-body integration. Such behaviour was also found by \citet{Moeckel12} when examining the subsequent dynamical evolution of SPH simulation of star formation by \citet{Bonnell08}. Furthermore,  in $N$-body models of clusters containing several hundred stars in virial equilibrium, 
\citet*{Gieles12} and \citet{Parker12d} showed that the median surface density decreases simply due to two-body relaxation.

\subsection{Bound/unbound stars}

We track the number of stars that remain bound as a function of time by calculating the kinetic and potential energies for each star. The potential energy of an individual star, $V_i$, is given by:
\begin{equation}
V_i = - \sum\limits_{i \not= j} \frac{Gm_im_j}{r_{ij}},
\end{equation} 
where $m_i$ and $m_j$ are the masses of two stars and $r_{ij}$ is the distance between them. The kinetic energy of a star, $T_i$ is given thus:
\begin{equation}
T_i = \frac{1}{2}m_i|{\bf v}_i - {\bf v}_{\rm cl}|^2, 
\end{equation}
where ${\bf v}_i$ and ${\bf v}_{\rm cl}$ are the velocity vectors of the star and the centre of mass of the cluster, respectively. A star is bound if $T_i + V_i < 0$.

\begin{figure}
\begin{center}
\rotatebox{270}{\includegraphics[scale=0.4]{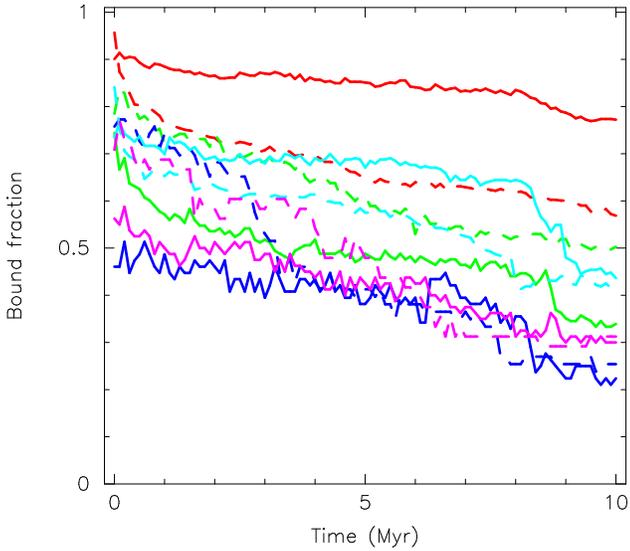}}
\end{center}
\caption[bf]{Evolution of the number fraction of bound stars. Simulations \emph{with} feedback are shown by the solid lines; simulations with no 
feedback are shown by the dashed lines. The red lines are Runs J, green lines are Runs I, dark blue lines are Runs UF, cyan lines are Runs UP and the magenta lines are Runs UQ. 
There is no correlation between the evolution fraction of bound stars and the initial conditions (i.e.\,\,feedback versus no feedback).}
\label{bound_3sims}
\end{figure}

In Fig.~\ref{bound_3sims} we show the number fraction of stars that remain bound over the $N$-body integration for all five pairs of simulations. Again, the simulations that formed with feedback are 
shown by the solid lines, and the simulations that formed without feedback are shown by the dashed lines. The colours correspond to the following simulations; red--Run~J, green--Run~I, dark blue--Run~UF, cyan--Run~UP and magenta--Run~UQ. 

Firstly, we note that the fraction of stars that are bound at the start of the $N$-body integration is less than 1 in all of the simulations, even those that are initially virialised, or slightly subvirial. This is due to the initially non-spherical, substructured 
nature of the simulations, which makes the definition of the centre of mass of the cluster problematic, and also that the individual velocities of stars which form on the outskirts of the cluster can be high. 

The simulations that formed from unbound clouds (UF, UP and UQ) have initially lower fractions of bound stars than the simulations that formed from bound clouds (Runs~J and I). 

The fraction of stars that are still bound at the instant of the removal of remaining gas appears to be slightly dependent on whether the cluster forms with or without feedback. In all of the simulation pairs, the cluster that formed with feedback has a lower fraction of bound stars initially than the cluster that formed without 
feedback (the solid and dashed lines in Fig.~\ref{bound_3sims}, respectively). However, in Runs J, I and UQ (the red, green and cyan lines) the cluster that formed with feedback retains more bound stars over the 10\,Myr of $N$-body integration. The remaining two simulations (UF and UQ) have similar final fractions of bound 
stars, irrespective of whether the clusters formed with or without feedback. 

\subsection{Mass segregation}

We use two different algorithms to search for mass segregation in the pairs of simulations. Firstly, we apply the Minimum Spanning Tree (MST) method of \citet{Allison09a}, which compares the MSTs of random sets of stars with the MST of a chosen subset. The ratio of the average random MST to the subset, 
$\Lambda_{\rm MSR}$, is significantly larger than unity if mass segregation is present \citep[such as is observed in the Orion Nebula Cluster, ][]{Hillenbrand98,Allison09a}, or less than unity if the most massive stars are more spread out than the average stars. The second algorithm is the $m - \Sigma$ 
method of \citet{Maschberger11}, which plots the surface density, $\Sigma$, of stars against their mass, $m$. Mass segregation is defined as the most massive stars having significantly larger local stellar surface densities compared to the average stars in the cluster.

There are two approaches to measuring mass segregation in clusters. One can either measure the global level of segregation \citep[i.e.\,\,including \emph{all} stars in the cluster -- e.g.][]{Allison09a,Olczak11,Parker11b}, or alternatively, identify subclusters and measure mass segregation 
on localised scales \citep[e.g.][]{Kirk10,Maschberger11,Girichidis12}. Here, we focus on global mass segregation, because we expect the subsequent $N$-body evolution to facilitate two-body relaxation and potentially lead to dynamical mass segregation when the stars are well-mixed.

We search for mass segregation at 0\,Myr, 1\,Myr and 5\,Myr in the simulations. Several clusters display hints of mass segregation, but not on a significant level. We therefore conclude that none of the clusters exhibit primordial, or subsequent dynamical mass segregation. 

\section{Discussion}

The main result in Section~\ref{results} can be summarised as follows. The evolution of cluster structure, as determined by the $\mathcal{Q}$--parameter, demonstrates that clusters which form under the influence of feedback retain primordial structure for longer than clusters that form 
without feedback in the simulations of \citet{Dale12a,Dale13}. Furthermore, three clusters that form with feedback remain substructured throughout the subsequent $N$-body evolution. This result in itself is not particularly helpful, as the differences in the amount of substructure can be merely relative; in two pairs of simulations 
(Runs~J and UP) both clusters attain a centrally concentrated morphology, but the clusters that formed with feedback have lower $\mathcal{Q}$--parameters. 

Furthermore, the evolution of the $\mathcal{Q}$--parameter can be highly stochastic. \citet{Parker12d} demonstrate that an initially highly substructured ($\mathcal{Q} < 0.5$), subvirial cluster ($T/|V| = 0.3$, where $T$ and $V$ are the total kinetic and potential energies of the stars in a cluster) 
erases its substructure within 0.5\,Myr as it becomes centrally concentrated due to violent relaxation \citep{Allison10}. An initially substructured and supervirial cluster ($T/|V| = 1.5$) expands and generally retains its substructure, although 1 in 10 clusters (with statistically identical initial conditions) 
become centrally concentrated due to the stochastic nature of the dynamical evolution \citep{Parker12d}. In both evolutionary scenarios, the range in final $\mathcal{Q}$--parameters between 10 identical simulations is $\sim 0.5$, implying that the differences between the simulations 
with and without feedback presented here could just be due to the stochastic dynamical evolution of the clusters over 10\,Myr.

However, there are several differences between the simulations in \citet{Allison10} and \citet{Parker12d} and the clusters we dynamically evolve here. Firstly, the fractal simulations in those papers assume highly correlated velocities on local scales \citep{Goodwin04a}, which increases the amount 
of dynamical mixing that takes place as the substructure violently relaxes. Secondly, those simulations were of fractal clusters with $N = 1000$ stars and radii of $\sim$1\,pc, (with a typical central density of $\sim$\,2000\,M$_\odot$\,pc$^{-3}$ following the subvirial collapse) whereas the most compact SPH simulation from \citet{Dale12a} and \citet{Dale13} is the cluster in Run~J that formed 
without feedback, which formed 578 stars and has a radius of $\sim$1\,pc (and central density of 220\,M$_\odot$\,pc$^{-3}$). At the other end of the scale, Run~UQ formed only 80 stars and has a radius of $\sim$12\,pc (and a central density of 0.6\,M$_\odot$\,pc$^{-3}$). 

Therefore, we expect the crossing time (and the related relaxation time) to be much longer for the simulations presented here. As an example, we compute the relaxation time, $t_{\rm relax}$, of a typical simulation from \citet{Allison10}, using the following formula \citep{Binney87}:
\begin{equation}
t_{\rm relax} = \frac{N}{8\,{\rm ln}\,N}t_{\rm cross},
\end{equation}
where $N$ is the number of stars and $t_{\rm cross}$ is the crossing time. For a typical simulation in \citet{Allison10}, $t_{\rm cross} = 0.1$\,Myr and $N = 1000$, which corresponds to a relaxation time $t_{\rm relax} = 2$\,Myr. On the other hand, the crossing time for the cluster in Run~J that formed with feedback 
is $t_{\rm cross} \sim 0.7$\,Myr, and its relaxation time (for $N = 685$ stars) is 9\,Myr. Conversely, the cluster in Run~J that  formed without feedback has a smaller radius (but similar velocity dispersion) and hence a much shorter crossing time ($t_{\rm cross} \sim 0.25$\,Myr), and a relaxation time of 2.8\,Myr. 
This is evident in the behaviour of the $\mathcal{Q}$--parameter; in Fig.~\ref{Qpar_all} the cluster in Run~J that formed without feedback (the red dashed line) has a constant $\mathcal{Q}$--parameter after $\sim$1.5\,Myr, whereas the simulation that formed with feedback (the red solid line) has a 
$\mathcal{Q}$--parameter that is still increasing at 10\,Myr.

A similar dependence of the evolution of the $\mathcal{Q}$--parameter on the relaxation time is found for each pair of simulations; in all cases, the clusters that formed with feedback have longer relaxation times, which is likely due to the initial densities of the cluster{\bf s}. In Fig.~\ref{Dens_all} 
we show that the initial volume density is lower for the simulations that form with feedback, due to gas expulsion, regulation of stellar mass growth, and the triggering and redistribution of star formation. We suggest that the two clusters that form with feedback which never erase their substructure are simply too diffuse to do so (Runs~UF~and~UQ,   
which have initial densities less than 1\,M$_\odot$\,pc$^{-3}$). This is also echoed by the initial surface densities of these clusters, which are 0.6\,stars\,pc$^{-2}$ for Run UF with feedback and 2\,stars\,pc$^{-2}$ for Run UQ with feedback. It is not possible to infer the initial density 
after 1\,Myr of dynamical evolution as the clusters all expand and relax, and clusters with very different initial densities evolve to have very similar densities (Figs.~\ref{Dens_all}, \ref{sigma_1} and \ref{sigma_2}).

Whilst it is possible to infer differences between the simulations with and without feedback, one needs to make an estimate of the initial density in order to calibrate the evolution of the $\mathcal{Q}$--parameter. As we have seen, a $\mathcal{Q}$--parameter which indicates 
a centrally concentrated morphology at an age of $>$\,2\,Myr could have formed with or without feedback. The only constraint we can place is that \emph{all} the clusters that formed \emph{without} feedback in the simulations of \citet{Dale12a} and \citet{Dale13} become centrally concentrated. 
If a cluster is observed at $\sim$5\,Myr to be substructured, it is either highly supervirial \citep{Parker12d}, and/or formed under the influence of feedback.

Other diagnostics of dynamical evolution do not distinguish between feedback and non-feedback star formation. Initial differences in the IMF (which is more top-heavy for simulations which are not regulated by feedback) are reduced by subsequent stellar evolution. Additionally, it is not 
possible to distinguish the two types of star formation by considering the kinematics of the clusters' evolution, as there is no systematic dependence on the fraction of bound stars.

\subsection{Scaled velocities}

\begin{figure}
\begin{center}
\rotatebox{270}{\includegraphics[scale=0.4]{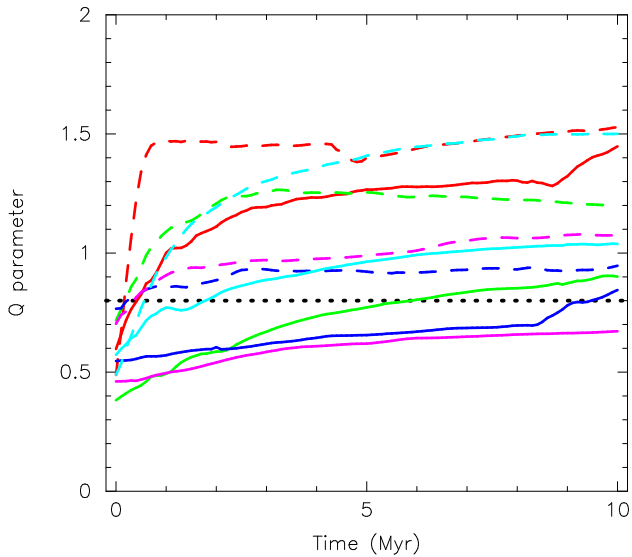}}
\end{center}
\caption[bf]{Evolution of the $\mathcal{Q}$-parameter for all simulations in which we scaled the velocities to be initially in virial equilibrium. The boundary between centrally concentrated, radially smooth clusters ($\mathcal{Q} > 0.8$) and substructured clusters ($\mathcal{Q} < 0.8$) is shown by the dotted line. 
Simulations \emph{with} feedback are shown by the solid lines; simulations with no feedback are shown by the dashed lines. The red lines are Runs J, green lines are Runs I, dark blue lines are Runs UF, cyan lines are Runs UP and the magenta lines are Runs UQ. The trend in the evolution of substructure is similar to the simulations with 
non-scaled velocities; however, the spread in $\mathcal{Q}$ between simulations is lower, and Run~I with feedback (the solid green line) loses its substructure earlier than in the non-scaled velocity case (at 5\,Myr instead of 8\,Myr).}
\label{Qpar_all_vir}
\end{figure}

In addition to our default initial conditions, we also performed a set of simulations in which we scaled the initial bulk motion of the stars in each cluster to be in virial equilibrium. Qualitatively, the results do not depend on the initial velocity set-up. There is again no trend in 
the evolution of the half-mass radius, bound fraction, surface density,  depending on whether the cluster formed with or without feedback. We show the evolution of structure, as measured by the $\mathcal{Q}$-parameter, for these (initially) virialised clusters in Fig.~\ref{Qpar_all_vir}. 
As in the case of the simulations with non-scaled velocities, the clusters that form with feedback retain more structure during the subsequent dynamical evolution.

Comparison of Fig.~\ref{Qpar_all_vir} to Fig.~\ref{Qpar_all} shows that the different velocity scaling has not made much difference to the evolution of the $\mathcal{Q}$--parameter. The simulation pair with the most stars (Run~J -- the red lines) does not become as centrally concentrated 
when in virial equilibrium as the unscaled simulations in Fig.~\ref{Qpar_all}, which may be due to the slightly subvirial initial velocities in the non-scaled simulations \citep[as subvirial clusters reach higher $\mathcal{Q}$--parameters,][]{Parker12d}. The remaining clusters 
follow similar evolution to those in the simulations that have the stellar velocities scaled to be in virial equilibrium, apart from Run~I with feedback (the solid green line) which reaches the border between structured and centrally concentrated earlier when in virial equilibrium 
(at 5\,Myr instead of 8\,Myr).

All other parameters (density evolution, fraction of bound stars) are similarly unchanged when the velocities of stars are scaled to be in virial equilibrium.

\section{Conclusions}

We have taken the end states of five pairs of hydrodynamical simulations \citep{Dale12a,Dale13}  and evolved them as $N$-body simulations for 10\,Myr 
to look for differences between the clusters in the simulations which formed under the influence and feedback, compared to the evolution of clusters that formed 
without feedback.

Gas expulsion and the other effects of photionization feedback -- the triggering, aborting and redistribution of star formation -- are expected to affect the initial dynamical states of young clusters, since they influence the masses, positions and velocities of the stars. The five pairs of $N$-body simulations presented here demonstrate clearly that this is the case, by comparing the subsequent evolution of clusters which either have or have not experienced photoionization feedback before supernovae are supposed to instantaneously clear them of gas.

Although the evolution of the feedback and no--feedback members of each pair of simulations is different, it is difficult to identify general trends in these differences -- for example, some feedback-influenced clusters lose stars more rapidly than their counterparts during their $N$-body evolution and some do not.

The only general result that we have been able to identify is that the feedback-influenced clusters which formed in the simulations of \citet{Dale12a,Dale13} tend to lose their substructure \citep[as measured by the $\mathcal{Q}$--parameter,][]{Cartwright04,Cartwright09} more slowly than their non-feedback--influenced opposite numbers. The most strongly feedback-affected simulations -- Runs I, UQ and UF from \citet{Dale12a} and \citet{Dale13} -- retain some substructure for 5--10 Myr, whilst none of the non-feedback simulations are able to support substructure for more than 1\,Myr of $N$-body evolution.

This difference is a straightforward consequence of the feedback clusters having lower volume densities and consequently longer relaxation times, so that in simulations UF and UQ, there is in fact little stellar mixing at all. These lower volume densities are in turn due to prolonged gradual mass loss from the parent clouds, and to the triggering and redistribution of star formation outside the principal subclusters.

However, the feedback and non-feedback clusters have varying degrees of \emph{initial} substructure, and there is no clear trend for the feedback calculations to have more or less initial substructure than their counterparts. It is thus very difficult to draw reliable conclusions about the influence feedback has had on a particular system from the observed $\mathcal{Q}$--parameter at any epoch. The strongest statement that could be made is that a substructured cluster $>$\,5\,Myr in age is likely to have experienced strong feedback. However, even this inference requires knowledge of the initial density of the cluster, and its initial virial state, and is based on a small number of simulations of small-$N$ clusters.\\

\section*{Acknowledgments}

We thank Diederik Kruijssen and Christoph Olczak for helpful discussions during the writing of this paper. The $N$-body simulations in this work were performed on the \texttt{BRUTUS} computing cluster at ETH Z{\"u}rich.

\bibliography{general_ref}

\label{lastpage}

\end{document}